\documentclass[pdflatex,sn-mathphys-num]{sn-jnl}% Math and Physical Sciences Numbered Reference Style
\usepackage{graphicx}%
\usepackage{multirow}%
\usepackage{amsmath,amssymb,amsfonts}%
\usepackage{amsthm}%
\usepackage{mathrsfs}%
\usepackage[title]{appendix}%
\usepackage{xcolor}%
\usepackage{textcomp}%
\usepackage{manyfoot}%
\usepackage{booktabs}%
\usepackage{algorithm}%
\usepackage{algorithmicx}%
\usepackage{algpseudocode}%
\usepackage{listings}%
\usepackage{tabularx}
\usepackage[version=4]{mhchem}
\usepackage{subcaption}
\usepackage{float}

\theoremstyle{thmstyleone}%
\theoremstyle{thmstyletwo}%

\theoremstyle{thmstylethree}%

\begin{document}

\title[Article Title]{An SO(3)-equivariant reciprocal-space neural potential for long-range interactions}

\author[1,3]{\fnm{Lingfeng} \sur{Zhang}}
\equalcont{These authors contributed equally to this work.}
\author[1]{\fnm{Taoyong} \sur{Cui}}
\equalcont{These authors contributed equally to this work.}
\author[2]{\fnm{Dongzhan} \sur{Zhou}}
\author[2]{\fnm{Lei} \sur{Bai}}

\author[2]{\fnm{Shufei} \sur{Zhang}}
\author*[3]{\fnm{Luca} \sur{Rossi}}\email{luca.rossi@polyu.edu.hk}
\author*[2,4]{\fnm{Mao} \sur{Su}}\email{sumao@pjlab.org.cn}

\author*[1,2]{\fnm{Wanli} \sur{Ouyang}}\email{wlouyang@ie.cuhk.edu.hk}

\author[1]{\fnm{Pheng-Ann} \sur{Heng}}

\affil[1]{\orgname{The  Chinese University of Hong Kong}, \orgaddress{\city{Hong Kong}, \postcode{999077}, \country{China}}}

\affil[2]{\orgname{Shanghai Artificial Intelligence Laboratory}, \orgaddress{\city{Shanghai}, \postcode{200232}, \country{China}}}

\affil[3]{\orgdiv{Department of Electrical and Electronic Engineering}, \orgname{The Hong Kong Polytechnic University}, \orgaddress{\city{Hong Kong}, \postcode{999077}, \country{China}}}

\affil[4]{\orgname{Shenzhen Institute of Advanced Technology, Chinese Academy of Sciences}, \orgaddress{\city{Shenzhen}, \postcode{518055}, \country{China}}}


\abstract{
Long-range electrostatic and polarization interactions play a central role in molecular and condensed-phase systems, yet remain fundamentally incompatible with locality-based machine-learning interatomic potentials. Although modern $SO(3)$-equivariant neural potentials achieve high accuracy for short-range chemistry, they cannot represent the anisotropic, slowly decaying multipolar correlations governing realistic materials, while existing long-range extensions either break $SO(3)$ equivariance or fail to maintain energy–force consistency. Here we introduce EquiEwald, a unified neural interatomic potential that embeds an Ewald-inspired reciprocal-space formulation within an irreducible $SO(3)$-equivariant framework. By performing equivariant message passing in reciprocal space through learned equivariant k-space filters and an equivariant inverse transform, EquiEwald captures anisotropic, tensorial long-range correlations without sacrificing physical consistency.
Across periodic and aperiodic benchmarks, EquiEwald captures long-range electrostatic behavior consistent with ab initio reference data and consistently improves energy and force accuracy, data efficiency, and long-range extrapolation. These results establish EquiEwald as a physically principled paradigm for long-range–capable machine-learning interatomic potentials.}




\maketitle
% \begin{figure}[h!] 
%   \centering 
%   \includegraphics[width=1.0\textwidth, trim=6cm 0 0 0, clip]{main.pdf}
%   \caption{Main} 
%   \label{fig:mylabel} 
% \end{figure}
\section{Introduction}\label{sec1}
Molecular dynamics (MD) simulation has become an indispensable tool for probing atomistic mechanisms underlying chemical reactivity, condensed-phase organization, materials discovery, and biomolecular function~\citep{hospital2015molecular, senftle2016reaxff, karplus1990molecular, yao2022applying, zitnick2022spherical}. Yet despite decades of progress, achieving quantum-level accuracy at the spatial and temporal scales required for realistic materials and molecular systems remains a central challenge. Quantum-mechanical methods such as density functional theory (DFT) deliver reliable energies and
forces~\cite{geerlings2003conceptual}, but their steep computational cost restricts system sizes and timescales. Classical force fields, while efficient, often lack the transferability and predictive fidelity needed for complex chemistries~\cite{deringer2021gaussian}. Machine-learning interatomic potentials (MLIPs) aim to overcome this trade-off by learning high-dimensional
potential energy surfaces from \textit{ab initio} data~\citep{ramprasad2017machine, butler2018machine, gubernatis2018machine, unke2021machine}, enabling simulations that approach quantum accuracy with near–force-field efficiency~\cite{carleo2019machine}.

Within this landscape, SO(3)-equivariant graph neural networks, including NequIP~\cite{batzner20223}, Allegro~\cite{musaelian2023learning}, and MACE~\cite{batatia2022mace}, have emerged as state-of-the-art short-range MLIPs. By explicitly encoding rotational and translational symmetries, these models achieve remarkable data efficiency and robustness across chemical space. However, their accuracy fundamentally relies on a strict locality assumption: the total energy is decomposed into atomic contributions determined solely by environments within a finite cutoff radius~\citep{anstine2023machine, schutt2018schnet, gasteiger2020directional}. This assumption is intrinsically incompatible with systems dominated by long-range interactions, such as electrostatics, dipole–dipole coupling, and collective polarization, which decay slowly and exhibit pronounced anisotropy~\citep{levin2009polarizable, bedrov2019molecular, borodin2009polarizable}. As a result, even the most advanced equivariant architectures struggle to represent the tensorial, multipolar correlations governing electrolytes, molecular crystals, and interfacial systems~\citep{anstine2023machine,cheng2025latent, gao2025foundation}. This limitation is not merely quantitative but representational: long-range physics cannot be faithfully captured by truncating interactions in real space, regardless of local model expressivity~\cite{huguenin2023physics}. At its core, this reflects a mismatch between the tensorial symmetry of long-range physical interactions and the truncated or scalar representations imposed by locality-based learning.

Efforts to incorporate long-range interactions into MLIPs have continued for years. Early work augments short-range models with empirical electrostatics and dispersion baselines, which enforce the correct asymptotic behavior but are typically system-dependent and hard to obtain for complex, heterogeneous environments~\citep{niblett2021learning, unke2021spookynet}. A more general line of work learns intermediate physical surrogates, most commonly effective partial charges or electronic proxies such as Wannier-center-based representations, and then evaluates long-range electrostatics explicitly~\citep{unke2019physnet,ko2021fourth,gao2022self, shaidu2024incorporating, zhang2022deep}. These approaches are easier to interpret and can capture polarization effects, but they often require additional supervision and remain sensitive to the non-uniqueness of charge partitioning. Message-passing models~\citep{schutt2017schnet,batzner20223,deng2023chgnet} extend the receptive field by stacking local convolutions, yet truly nonlocal couplings, such as interactions between fragments separated beyond the cutoff, are still difficult to capture. Inspired by Ewald summation, more recent work models long-range effects in reciprocal space by aggregating global interactions through Fourier components with learnable frequency filters~\citep{yu2022capturing, kosmala2023ewald, king2025machine}. However, these frameworks encode reciprocal-space signals using scalar structure factors or latent charge-like variables, which average over angular information and can limit the description of directional electrostatics, multipolar correlations, and anisotropic polarization. LODE methods~\citep{huguenin2023physics, grisafi2019incorporating} provide a complementary strategy by using local descriptors to parameterize Coulomb and more general $1/r^{p}$ far-field interactions. This formulation preserves analytic asymptotic behavior and enables the incorporation of physically motivated long-range terms within equivariant frameworks. However, because the long-range contributions are introduced through predefined functional forms, these approaches can be less flexible in capturing complex environment-dependent electronic correlations, and they do not fully integrate long-range interactions into a unified end-to-end learning architecture.

Here, we introduce \textbf{EquiEwald}, a unified SO(3)-equivariant neural interatomic potential that embeds Ewald summation within an irreducible representation (irrep) framework~\cite{duval2023hitchhiker}. EquiEwald replaces the conventional locality assumption with a degree-resolved reciprocal-space pathway that enables long-range interactions to be represented at the tensorial level. By performing $SO(3)$-equivariant message passing directly in reciprocal space, the model can capture anisotropic and non-local correlations while preserving exact rotational symmetry and energy–force consistency. Specifically, EquiEwald computes irrep-resolved structure factors, applies learned $k$-space filters under both periodic and aperiodic boundary conditions, and maps spectral signals back to atomic space through an equivariant inverse transform, yielding a unified treatment of real-space and reciprocal-space physics within a single differentiable architecture. Across a suite of demanding benchmarks, we show that this representation-level integration of reciprocal-space information leads to improved accuracy, data efficiency, and transferability in systems governed by non-local interactions. While EquiEwald does not aim to replace explicit electrostatic models, analytic Ewald formulations, or environment-dependent dielectric descriptions, it provides a physically structured, equivariant alternative for learning long-range interactions directly from \textit{ab initio} data, addressing a fundamental representational limitation of locality-based MLIPs.

% \end{itemize}
\section{Results}\label{sec2}
\subsection{Preliminaries}

MLIPs aim to learn mappings from atomic structures to physical observables such as total energy $E$ and forces $\mathbf{F}_i = -\nabla_{\mathbf{r}_i} E$.
In the absence of external fields, the underlying physical laws are invariant under global translations, rotations, and permutations of atoms with identical species.
To ensure physical consistency and improve generalization, modern neural potentials typically enforce \textit{equivariance} to these symmetries~\cite{batzner20223, musaelian2023learning}.
% $\ell$
In this work, we focus on $SO(3)$-equivariance, which requires that the learned atomic features transform as irreducible spherical tensors under global 3D rotation~\citep{thomas2018tensor, geiger2022e3nn}.
Formally, for a rotation matrix $R \in SO(3)$ and atomic positions $\{\mathbf{r}_i\}_{i=1}^N$, an $\ell$-type atomic representation $\mathbf{x}_i^{(\ell)} \in \mathbb{C}^{2\ell+1}$ must satisfy
\begin{equation}
\mathbf{x}_i^{(\ell)} \mapsto \mathbf{x}_i^{(\ell)\prime} = D^{(\ell)}(R) \, \mathbf{x}_i^{(\ell)},
\end{equation}
where $D^{(\ell)}(R)$ is the Wigner-$D$ matrix of degree $\ell$~\citep{weiler20183d}. A formal proof demonstrating the $SO(3)$ equivariance of each step within the EquiEwald module is provided in Supplementary Section 2. This property ensures that energy predictions remain invariant ($E' = E$) and that forces transform as proper vectors ($\mathbf{F}_i' = R \mathbf{F}_i$) under arbitrary global rotation~\citep{chmiela2017machine}.

Throughout this paper, we use this formulation to build rotation-equivariant neural architectures.
Our proposed long-range module, EquiEwald, operates on such irreducible $SO(3)$ representations in reciprocal space and preserves equivariance via spherical harmonic decomposition, frequency-domain filtering, and inverse accumulation, as detailed in the following sections.

\begin{figure}[t]
  \centering
  \includegraphics[width=\textwidth,trim=6cm 0 0 0,clip]{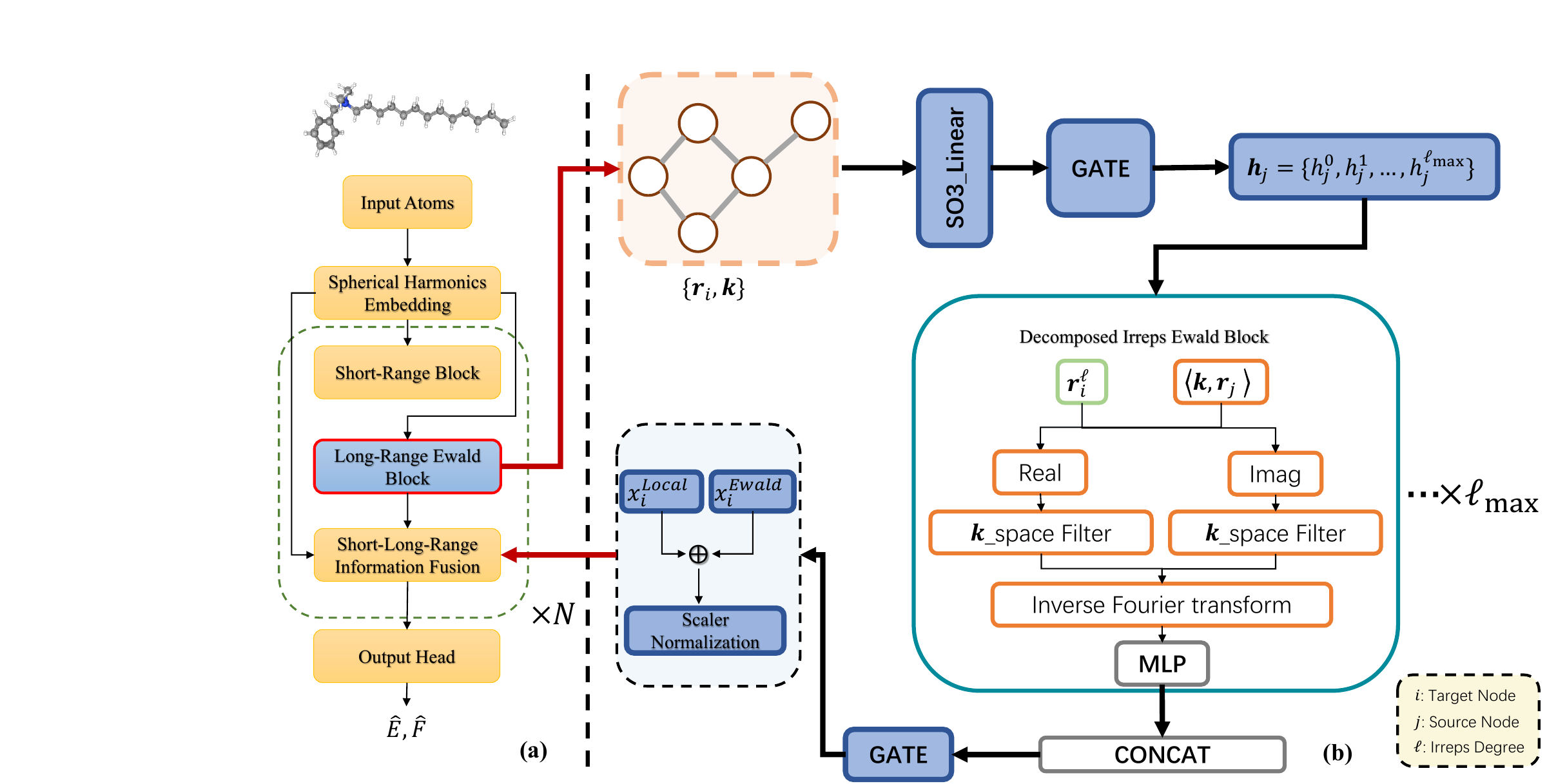}
    \caption{\textbf{Overview.} \textbf{a}, The whole model structure of EquiEwald. \textbf{b}, The EquiEwald long-range block takes node features and wave-vector inputs, applies an $\mathrm{SO}(3)$ linear map and gating, then decomposes degree-$\ell$ irreps into real/imag branches with $k$-space filters driven by $\langle \mathbf{k}, \mathbf{r}_j\rangle$; an inverse Fourier transform and an MLP return real-space updates per degree. Outputs across degrees are concatenated and gated, then fused with normalized local features to yield a rotationally equivariant long-range interaction update.}
  \label{fig:main}
\end{figure}

\subsection{Overall structure of EquiEwald}

EquiEwald is a neural framework for learning interatomic potentials that unifies short-range geometric modeling with long-range physical interactions in a single, rotation-equivariant architecture. As shown in Fig.~\ref{fig:main}(a), it consists of two synergistic representation pathways: a short-range encoder that captures local atomic environments using a local graph message passing backbone, and a long-range spectral encoder that introduces non-local information via message passing in reciprocal space. Both components operate on the same atom-wise input and their outputs are fused through a residual update to form the final atomic representation used for energy and force prediction.

A central feature of the EquiEwald design, illustrated in Fig.~\ref{fig:main}(b), is its use of high harmonic spherical degrees to organize and propagate information in reciprocal space. This enables the model to resolve anisotropic, direction-dependent, and multipolar correlations that are critical for accurately modeling long-range interactions in systems with electrostatics, polarization, or delocalized electronic structure. Unlike purely local models that rely on cutoff-based neighborhoods, EquiEwald captures extended spatial dependencies through equivariant Fourier accumulation and degree-resolved filtering, maintaining $SO(3)$ symmetry throughout.

By embedding physically motivated priors into the representation space, EquiEwald provides a unified treatment of short-range and long-range physics effects. This design improves model accuracy, long-range extrapolation, and data efficiency across both periodic and aperiodic systems, including charged dimers, conjugated molecules, supramolecular assemblies, and biomolecular dynamics.

\subsection{Experimental results}
The effectiveness of EquiEwald was systematically evaluated on both aperiodic and periodic systems. For the main results, we adopted the eSCN framework~\cite{passaro2023reducing} as our primary method. For periodic systems, we additionally included EquiformerV2~\cite{liao2023equiformerv2} as another baseline. Aperiodic benchmarks included molecular dimer~\cite{king2025machine}, AIMD-Chig~\cite{wang2023aimd}, and buckyball catcher~\cite{chmiela2023accurate} datasets, while periodic systems were assessed using the OC20 dataset~\cite{chanussot2021open}. 
% Data conformations are provided in Supplementary Fig. S1.

% Periodic systems were assessed using the OC20 dataset~\cite{chanussot2021open}, while aperiodic benchmarks included molecular dimer~\cite{king2025machine}, AIMD-Chig~\cite{wang2023aimd}, and buckyball catcher~\cite{chmiela2023accurate}datasets. Data conformations are provided in Supplementary Fig. S1. Additional results for the EquiformerV2~\cite{liao2023equiformerv2} framework on the OC20 dataset are reported in Supplementary Section 3. 

\begin{figure}[H]
  \centering
  % \makebox[\textwidth]{ 
  %   \begin{subfigure}[t]{0.4\textwidth}
  %     \centering
  %     \includegraphics[width=\textwidth]{Image/Fig2_cumulene.pdf}
  %     \captionsetup{labelformat=empty}
  %     \label{fig:cumulene_fig2}
  %   \end{subfigure}
  % }

  % \makebox[\textwidth]{ 
  %   \begin{subfigure}[t]{0.6\textwidth}
  %     \centering
  %     \includegraphics[width=\textwidth]{Image/Cumulene_main_new.pdf}
  %     \captionsetup{labelformat=empty}
  %     \caption{\textbf{Cumulene}}
  %     \label{fig:cumulene}
  %   \end{subfigure}
  % }

  \makebox[\textwidth]{ 
    \begin{subfigure}[t]{0.8\textwidth}
      \centering
      \includegraphics[width=\textwidth]{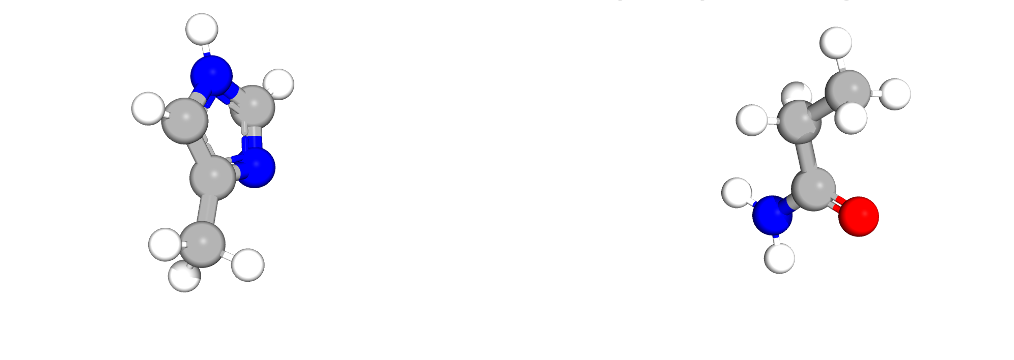}
      \captionsetup{labelformat=empty}
      \label{fig:dimer_fig2}
    \end{subfigure}
  }

  \makebox[\textwidth]{
    \begin{subfigure}[t]{1.0\textwidth}
      \centering
      \includegraphics[width=\textwidth]{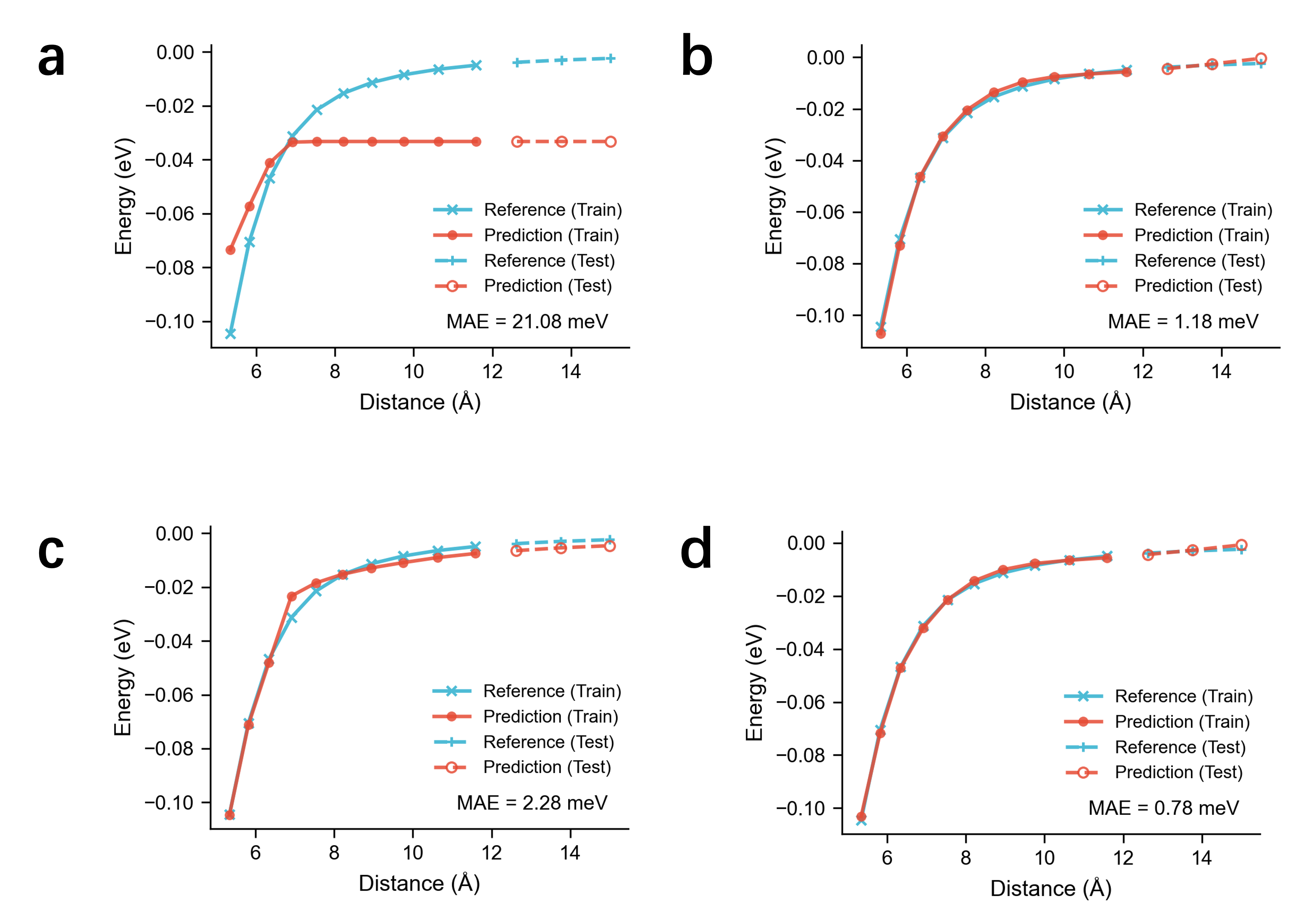}
      \captionsetup{labelformat=empty}
      % \caption{\textbf{Charged molecular Dimer}}
      \label{fig:dimer}
    \end{subfigure}
  }
  
    \caption{\textbf{Benchmark comparison between short-range and long-range models on charged molecular dimer systems.} 
    \textbf{a--d} Predicted versus reference energy results for polar-polar dimers. The short-range model is compared with three long-range variants: (\textbf{b}) eSCN+EwaldMP (scalar)~\cite{kosmala2023ewald}, (\textbf{c}) eSCN+LES~\cite{cheng2025latent}, and (\textbf{d}) eSCN+EquiEwald. For energy prediction, the short-range model fits only short-distance configurations and fails to generalize beyond the cutoff, while all long-range variants recover the full interaction curve, including distant test points.}
  \label{fig:Dimer}
\end{figure}

\vspace{0.5em}
\noindent \textbf{Molecular Dimer.}
This system presents a prototypical case of long-range electrostatic interaction between spatially separated charged species. It comprises a cationic $\mathrm{C}_{4}\mathrm{N}_{2}\mathrm{H}_{6}$ and an anionic $\mathrm{C}_{3}\mathrm{N}\mathrm{O}\mathrm{H}_{7}$ molecule, forming a loosely bound dimer where intermolecular interactions are governed predominantly by long-range dipole-dipole forces. The training configurations cover centroid separations from 5 to 12~\AA, while the test set focuses on even more distant geometries extending from 12 to 15~\AA. At these distances, the two fragments lie well beyond the 5~\AA\ local cutoff, rendering them effectively disconnected in standard message-passing architectures. Figure~\ref{fig:Dimer} compares the performance of four modeling approaches. The baseline eSCN model (Fig.~\ref{fig:Dimer}a) does not capture the correct asymptotic decay of the interaction energy and yields a large MAE of 21.08 meV. As interactions beyond the local cutoff are not modeled, the predicted energies level off too early and break down in the extrapolative test set. Two alternative long-range methods are evaluated to address this issue (Fig.~\ref{fig:Dimer}b and c). These methods substantially reduce the error, achieving MAEs of 1.18 meV and 2.28 meV, respectively, but visible deviations from the reference values remain at larger separations. In contrast, the eSCN + EquiEwald model (Fig.~\ref{fig:Dimer}d) produces accurate energy predictions across the full range of separations, including the extrapolative test configurations, with a MAE of 0.78 meV. We also compared the results for another pair of molecular dimers. A detailed comparison of the energy and force MAE between eSCN and eSCN + EquiEwald is given in Supplementary Fig.~S1. These results show that EquiEwald overcomes the limitations of scalar methods in modeling long-range electrostatic interactions by incorporating higher-order tensor representations, enabling accurate capture of anisotropic and multipolar interactions, and significantly improving the modeling accuracy and physical consistency for molecular systems.

% the baseline eSCN model struggles to extrapolate beyond the short-range regime. It fails to capture the asymptotic decay of interaction energy and already shows substantial errors at the tail end of the training range. For test structures with large separations, the energy predictions collapse entirely. In contrast, the eSCN + EquiEwald model yields accurate predictions across the entire distance spectrum, including configurations in the extrapolative test regime. The effectiveness of our approach is also reflected in the force predictions, as shown in Fig.~\ref{fig:cumuleneAndDimer}d. The EquiEwald-enhanced model closely aligns with the ab-initio ground truth across all configurations, while the baseline model exhibits large, systematic deviations and often degenerates to near-zero force magnitudes. The explicit comparison of energy and force MAE between eSCN and eSCN + EquiEwald is detailed in Supplementary Table S1 and Fig. S2. These results demonstrate that EquiEwald can overcome the limitations of locality in message passing by enabling accurate modeling of long-range electrostatics in sparsely interacting molecular systems. 

\vspace{0.5em}
\noindent \textbf{AIMD-Chig.} Derived from ab initio molecular dynamics of the protein Chignolin, it poses challenges due to its flexible backbone, complex conformational dynamics, and long-range intra-molecular interactions important for maintaining structural stability. As shown in Table~\ref{tab:Chig}, the baseline eSCN model yields a test energy MAE of 193.9 meV and a force MAE of 23.1 meV/\AA. With EquiEwald, these errors are reduced to 109.0 meV and 18.1 meV/\AA, corresponding to improvements of 44\% and 21\%, respectively. By incorporating reciprocal-space information with angular resolution, EquiEwald provides the representational capacity needed to model collective motions and long-range electrostatics in biomolecular systems. Table~\ref{tab:Chig} further includes scalar EwaldMP, enabling a systematic comparison between invariant and equivariant treatments of long-range interactions.

% As shown in Supplementary Table S2, EquiEwald significantly outperforms a scalar Ewald message-passing baseline. 

% To demonstrate the advantage of equivariant long-range modeling, we compare EquiEwald with the EwaldMP baseline on the chignolin system. Chignolin exhibits a highly flexible backbone and is stabilized by long-range intra-molecular interactions that depend on relative orientations and collective motions. Non-equivariant long-range models represent such interactions using rotation-invariant scalar features, which average out directional information and limit their ability to distinguish distinct conformations. In contrast, EquiEwald preserves SO(3)-equivariance in reciprocal space, enabling long-range correlations to be modeled in an orientation-aware manner across the entire protein. The consistently superior performance observed in our experiments highlights the necessity of equivariant long-range representations for accurately capturing the conformational energy landscape of flexible biomolecules.
\begin{table}[t]
\caption{Performance comparison of different long-range interaction modeling approaches on the AIMD-Chig dataset.
The baseline eSCN model is compared with variants incorporating scalar EwaldMP, and the proposed EquiEwald method. Mean Absolute Errors (MAE) for energy and atomic forces are reported in \textbf{meV} and \textbf{meV/\AA}, respectively.}
\label{tab:Chig}
\centering
\begin{tabular}{lcc}
\toprule
Model & Test Energy (MAE) & Test Force (MAE) \\
\midrule
eSCN~\citep{passaro2023reducing} & 193.9 & 23.1 \\
% eSCN + CACE-lr & 193.8 & 22.7 \\ 
eSCN+EwaldMP (scalar)~\citep{kosmala2023ewald} & 132.8 & 20.3 \\
eSCN+EquiEwald & \textbf{109.0} & \textbf{18.1} \\
\bottomrule
\end{tabular}
\end{table}

% We further evaluate the effect of explicitly modeling long-range interactions on thermodynamic properties with experiments of various fast-folding Chignolin.

We further evaluate the effect of explicitly modeling long-range interactions on \textbf{thermodynamic properties} using the fast-folding protein Chignolin. From the simulation trajectories, $100{,}000$ conformational snapshots were uniformly sampled and analyzed using the $Q$ score, which measures the fraction of native contacts relative to the reference folded structure. The model-predicted potential energies were then used to reconstruct the energy landscape and estimate thermodynamic quantities, including the folding free-energy difference ($\Delta G$). Detailed protocols and analysis procedures are described in the Methods. Compared to the baseline eSCN model, which relies solely on local message passing, incorporating the EquiEwald reciprocal-space module substantially improves the accuracy of energy predictions relevant to free-energy estimation. Specifically, the prediction error is reduced from $1.15$ kcal/mol for eSCN to $0.67$ kcal/mol for eSCN+EquiEwald, corresponding to an approximate $42\%$ relative reduction. This improvement demonstrates that explicitly modeling long-range electrostatic interactions leads to more accurate thermodynamic estimates. In contrast, purely local representations struggle to capture the collective intra-molecular interactions that stabilize folded conformations in Chignolin.

% By embedding Ewald summation directly into an $SO(3)$-equivariant reciprocal-space representation, EquiEwald enables the model to resolve non-local and anisotropic correlations across the protein backbone and side chains.
% As a consequence, the enhanced model provides a more faithful description of the underlying potential energy landscape, leading to improved free-energy accuracy in biomolecular systems dominated by subtle long-range interactions.

\noindent \textbf{Buckyball Catcher.} The Buckyball Catcher system presents a strong challenge for interatomic potentials due to its supramolecular structure and long-range non-covalent interactions between the host and the encapsulated fullerene. These interactions exceed typical cutoff distances, making them difficult for local models to capture. As shown in Supplementary Table S1, the baseline eSCN model yields a test energy MAE of 36.0 meV, while incorporating EquiEwald reduces this to 18.1 meV, corresponding to an improvement of nearly 50\%. For force prediction, the MAE drops from 6.4 meV/\AA\ to 6.1 meV/\AA. This larger energy gain arises because reciprocal-space methods efficiently capture the global modes governing the potential energy surface, whereas local force gradients remain more sensitive to high-frequency spectral truncation. These improvements are consistently observed across multiple training runs, as evidenced by the error bars in Supplementary Fig.~S2, which demonstrate the statistical stability of our method. These results demonstrate that EquiEwald enables the model to capture extended spatial correlations through reciprocal-space message passing with high harmonic spherical degrees, improving accuracy in systems with long-range supramolecular interactions.

% \vspace{0.5em}
\noindent \textbf{OC20.}
We further evaluate EquiEwald on the OC20 S2EF benchmark, which comprises catalytic surface--adsorbate structures under periodic boundary conditions. Because OC20 contains heterogeneous interfaces, the predicted energies and forces can be affected by nonlocal effects, including charge redistribution, surface polarization, and interactions induced by periodicity. This property makes OC20 a relevant benchmark for evaluating long-range interaction modeling. As shown in Table~\ref{tab:OC20}, EquiEwald consistently improves the performance of both backbones. The improvement is particularly pronounced for EquiformerV2, for which the energy and force MAEs decrease from 541.0 to 453.0 meV and from 46.4 to 38.4 meV/\AA, respectively. For eSCN, EquiEwald also reduces the energy MAE from 347.0 to 321.2 meV and the force MAE from 24.7 to 24.1 meV/\AA. Overall, these results indicate that reciprocal-space long-range modeling is beneficial for periodic interfacial systems, with a more evident effect on energy prediction than on force prediction. 
% The reason is that long-range interactions contribute more directly to the global energy landscape, whereas their contributions to atomic forces may be partially cancelled through directional superposition in relatively periodic or locally isotropic environments.
% Based on the experimental results obtained in this study using the OC20 S2EF/2M training dataset and evaluated on four validation splits (Table~\ref{tab:OC20}), the eSCN and EquiformerV2 incorporating the EquiEwald method demonstrates improvements over the baseline eSCN model in both energy and force prediction tasks. This larger energy gain arises because reciprocal-space methods efficiently capture the global modes governing the potential energy surface, whereas local force gradients remain more sensitive to high-frequency spectral truncation. These improvements are attributed to the ability of the EquiEwald method to capture long-range interactions in a physically consistent and $SO(3)$-equivariant manner, which is particularly important for periodic systems such as those in OC20. In such systems, electrostatic and polarization effects arising from periodic boundary conditions significantly influence both total energies and atomic forces. By embedding the Ewald summation directly into the model's convolutional architecture and performing it on high harmonic spherical degrees, EquiEwald incorporates reciprocal-space information with fine angular resolution. This enables the model to represent anisotropic and multipolar correlations that are otherwise inaccessible to local or cutoff-based methods.

\begin{table}[t]
\caption{Mean absolute errors (MAE) of energy and force. Results are reported on the evaluation splits for OC20. Energy errors are reported in \textbf{meV}, and force errors are in \textbf{meV/\AA}. Our method is highlighted in bold.}
\label{tab:OC20}
\centering
\begin{tabular}{@{}lcc@{}}
\toprule
Model & Test(Evaluation) Energy & Test(Evaluation) Force \\
\midrule
eSCN~\citep{passaro2023reducing} & 347.0 & 24.7 \\
eSCN+EquiEwald & \textbf{321.2} & \textbf{24.1} \\
\midrule
EquiformerV2~\citep{liao2023equiformerv2} & 541.0 & 46.4 \\
EquiformerV2+EquiEwald & \textbf{453.0} & \textbf{38.4} \\
\bottomrule
\end{tabular}
\end{table}

\section{Discussion}
In this work, we introduced EquiEwald, an $SO(3)$-equivariant neural interatomic potential that integrates long-range interactions directly into the representation space of message-passing models. By embedding Ewald summation within a reciprocal-space framework and performing degree-resolved equivariant convolutions over high harmonic spherical components, EquiEwald captures multipolar and anisotropic correlations that are inaccessible to purely local methods. Our results across diverse benchmarks demonstrate the effectiveness and generality of this approach. In periodic systems, supramolecular assemblies, conjugated molecules, charged dimers, and biomolecular dynamics, EquiEwald consistently improves energy and force predictions, even under limited data. These improvements reflect the model’s ability to resolve long-range electronic, electrostatic, and structural dependencies in a physically consistent manner. Rather than treating long-range interactions as external corrections, EquiEwald unifies short-range chemistry and long-range physics within a single, differentiable neural potential. This reframes the challenge of long-range modeling in MLIPs as a representational problem, and highlights the importance of embedding physical structure into the architecture itself, rather than deferring it to post hoc corrections or auxiliary terms.

By bridging quantum accuracy and force-field scalability, EquiEwald offers a promising direction for next-generation MLIPs applicable to realistic systems such as electrolytes, molecular crystals, interfaces, and proteins. Its reciprocal-space formulation may also be extended to incorporate dielectric response, long-range polarization, or time-dependent interactions in future work. More broadly, it suggests a general strategy for integrating global physical priors into geometric deep learning models for scientific applications.

\section{Methods}
\subsection{Model implementations}
\noindent \textbf{Short-range Encoders.} 
To capture local geometric information, we employ equivariant neural networks that process atomic structures while maintaining physical symmetries. We initialize the initial atom features $\mathbf{x}_i^0$ via an embedding of the atomic number $z_i$. To represent the directional nature of the local relation, these encoders utilize spherical harmonics to expand the relative displacement vectors between neighboring atoms. Here, the spherical harmonic degree $\ell$ determines the resolution of the angular information, where higher degrees capture more complex multi-body correlations, and the magnetic component $m$ (ranging from $-\ell$ to $\ell$) tracks the specific orientation of these features under rotation. Through equivariant operations, the encoder iteratively updates atom features to produce the final local atom representation, denoted as $\mathbf{x}_i^{\text{local}}$. Specific architectural details (eSCN~\citep{passaro2023reducing} and EquiformerV2~\citep{liao2023equiformerv2}) are provided in the Supplementary Section 3.
% Short-range encoders are employed to capture local geometric information from the atomic structure. We initialize the atom feature $\mathbf{x}^{0}_{i}$ via an embedding of the atomic number $z_i$, and use the eSCN~\cite{passaro2023reducing} architecture as the backbone to validate our method.

% eSCN is designed to efficiently capture high-order angular information inherent to atomistic systems. It adopts a rotation-based reduction strategy, where edge vectors between atoms are aligned with a quantization axis by means of Wigner-D matrices. This procedure converts the computationally expensive dense $SO(3)$ spherical convolutions into sparser and more tractable $SO(2)$ operations, greatly enhancing efficiency within the message passing layers.
% eSCN naturally supports high spherical harmonic degrees and manages the mixed-symmetry channels required to represent anisotropic, direction-dependent features. Through its rotation-aligned convolutions and effective handling of high-degree tensors, eSCN is able to learn expressive representations of local geometry and angular correlations that are critical for predictive accuracy in quantum and molecular modeling tasks. We denote the final output of the short-range encoder as $\mathbf{x}_i^{\text{Local}}$.

\vspace{0.5em}
\noindent \textbf{Long-range Encoders.} 
The EquiEwald block is designed to capture non-local directional interactions by extending the Ewald message passing~\cite{kosmala2023ewald} framework to high-degree equivariant representations. While the original Ewald MP primarily communicates isotropic scalar information, our high-degree approach enables the model to learn anisotropic long-range correlations and complex multi-body geometric patterns that extend across the entire molecular system. By incorporating tensorial features ($\ell > 0$), the block captures how the specific orientations and angular distributions of distant atomic environments mutually influence the potential energy surface, effectively modeling long-range directional dependencies that are lost in local or scalar-only paradigms.

The core concept of EquiEwald involves flipping the traditional rationale behind Ewald summation, rather than starting with a known physical kernel and seeking a decomposition, we parametrize a learnable filter that is specifically not short-ranged. By learning this component directly in Fourier space, we replace the conventional spatial distance limit with a cutoff on frequency, enabling the efficient capture of global structural information through reciprocal-space message passing. To process these high-degree representations, EquiEwald introduces degree-resolved message passing in reciprocal space. For each spherical harmonic degree $\ell \in \{0, 1, \dots, \ell_{\max}\}$, we first compute structure factor embeddings $s_{\alpha,m}^{(\ell)}$ via a forward Fourier transform:
\begin{equation}
s_{\alpha,m}^{(\ell)}
= \sum_{j\in\mathcal{S}} \mathbf{x}_{j,m}^{(\ell)}
\exp\!\big(-i\,\mathbf{k}_\alpha \cdot \mathbf{r}_j\big)
D(\mathbf{k}_\alpha,\mathbf{r}_j),
\end{equation}
where $\mathbf r_j$ denotes the position of atom $j$, $\mathbf k_\alpha$ indexes the reciprocal-space sampling points, and $\mathbf x_{j,m}^{(\ell)}$ is the $m$-th magnetic component of the degree-$\ell$ atomic feature. The scalar factor $D(\mathbf k,\mathbf r)$ denotes an accumulation window that weights each phase term during reciprocal-space summation; in our implementation this window is applied symmetrically in both the forward and inverse accumulation. In periodic systems we set $D(\mathbf k,\mathbf r)=1$ since structure factors are evaluated directly on the reciprocal lattice. In aperiodic systems, where reciprocal space is discretized on a Cartesian voxel grid with spacing $\Delta k$, we use a voxel-averaging window to account for finite $k$-space resolution.
% where $\mathbf{r}_j$ denotes the position of atom $j$ , $\mathbf{k}_{\alpha}$ indexes the reciprocal-space sampling points, and $\mathbf{x}_{j,m}^{(\ell)}$ represents the $m$-th magnetic component of the degree-$\ell$ atomic feature. The scalar damping function $D(\mathbf{k}, \mathbf{r})$ is applied to ensure numerical stability and account for specific boundary conditions, such as voxel-averaging in aperiodic systems. In our implementation the voxel window is applied symmetrically in both the forward and inverse accumulation; we use the same notation $D(\mathbf k,\mathbf r)$ for both.

To maintain $SO(3)$ equivariance throughout the global update, a learnable filter $\mathbf{F}(\mathbf{k}_{\alpha})$ is applied identically to all magnetic components $m$ within a specific degree $\ell$ before the inverse transform:
\begin{equation}
M_{m}^{(\ell)}(\mathbf{r}_i)
= \sum_{\alpha=1}^{N_k}
\exp\!\big(i\,\mathbf{k}_\alpha \cdot \mathbf{r}_i\big)
\mathbf{F}(\mathbf{k}_\alpha)\,
s_{\alpha,m}^{(\ell)}.
\end{equation}
By constraining $\mathbf{F}(\mathbf{k}_{\alpha})$ to mix the channel dimensions, EquiEwald ensures the geometric integrity of the tensorial features is preserved. We use $\mathbf{F}(\mathbf{k}_\alpha)$ as a unified notation: in periodic systems it reduces to a shared, $k$-independent channel mixer $\mathbf{F}\in\mathbb{R}^{C\times C}$, whereas in aperiodic systems it denotes a $k$-dependent diagonal spectral gate $\mathbf{F}(\mathbf{k}_\alpha)=\mathrm{diag}(\mathbf{f}_\alpha)$ parameterized by the radial embedding $\psi(\|\mathbf{k}_\alpha\|)$. The resulting reciprocal-space message $M_{m}^{(\ell)}(\mathbf{r}_i)$ provides each atom with a comprehensive global update, complementing the local representations $\mathbf{x}_i^{\text{local}}$ to capture the full spectrum of interatomic interactions.

\vspace{0.5em}
\noindent \textbf{Periodic systems.} For periodic systems, reciprocal vectors are sampled on the reciprocal lattice induced by the simulation cell.
Let $(\mathbf a_1,\mathbf a_2,\mathbf a_3)\in\mathbb R^3$ be the direct lattice basis and
$(\mathbf b_1,\mathbf b_2,\mathbf b_3)$ the reciprocal basis satisfying
$\mathbf b_i\!\cdot\!\mathbf a_j = 2\pi\delta_{ij}$, i.e., with $V=\mathbf a_1\!\cdot\!(\mathbf a_2\times\mathbf a_3)$,
\begin{align}
\mathbf{b}_1 &= \tfrac{2\pi}{V}(\mathbf{a}_2\times\mathbf{a}_3),\quad
\mathbf{b}_2 = \tfrac{2\pi}{V}(\mathbf{a}_3\times\mathbf{a}_1),\quad
\mathbf{b}_3 = \tfrac{2\pi}{V}(\mathbf{a}_1\times\mathbf{a}_2).
\end{align}
% For periodic systems, reciprocal vectors $\mathbf{k}$ lie on the reciprocal lattice determined by the unit cell. We therefore evaluate structure factors directly at these lattice $\mathbf{k}$-points.
% Because the $\mathbf{k}$-set is obtained analytically from the cell geometry, no grid-interpolation or anti-aliasing window is required, and we use $D(\mathbf{k},\mathbf{r})=1$. The long-range update is implemented as a learnable spectral filter applied to the structure factors in reciprocal space. To preserve $SO(3)$-equivariance of tensor features, the filter will not mix magnetic components $m$ within a given degree $\ell$; it only mixes the channel dimension via a shared linear map $\mathbf{F}^{(\ell)}$, implemented in low-rank form $\mathbf{F}^{(\ell)}=\mathbf{W}_{\mathrm{up}}\mathbf{W}_{\mathrm{down}}$.
Following Ewald message passing, we truncate reciprocal sampling by an index box rather than a direct radial cutoff to keep the number of $\mathbf{k}$-points fixed across structures. Specifically, we consider integer triples
\begin{equation}
(n_x,n_y,n_z)\in\mathcal I
=\{-N_x,\ldots,N_x\}\times\{-N_y,\ldots,N_y\}\times\{-N_z,\ldots,N_z\},
\label{eq:pbc_index_box}
\end{equation}
and form reciprocal vectors
\begin{equation}
\mathbf{k}_{n_x,n_y,n_z}=n_x\mathbf b_1+n_y\mathbf b_2+n_z\mathbf b_3.
\label{eq:pbc_k_def}
\end{equation}
We then enumerate the resulting set $\{\mathbf{k}_{n_x,n_y,n_z}:(n_x,n_y,n_z)\in\mathcal I\}$ and relabel it as
$\{\mathbf{k}_\alpha\}_{\alpha=1}^{N_k}$ for convenient summation, where each $\alpha$ corresponds to one triple in $\mathcal I$.

Since $\{\mathbf{k}_\alpha\}_{\alpha=1}^{N_k}$ are reciprocal-lattice vectors computed analytically from the unit-cell geometry, the Fourier accumulation is performed directly on this lattice without any interpolation.
Accordingly, no additional sampling window is required and we set $D(\mathbf{k},\mathbf{r}) = 1$. Moreover, under a global rotation $R\in SO(3)$, the crystal lattice co-rotates: $\mathbf{a}_i\mapsto R\mathbf{a}_i$ implies $\mathbf{b}_i\mapsto R\mathbf{b}_i$ and hence $\mathbf{k}_\alpha\mapsto R\mathbf{k}_\alpha$. Together with $\mathbf{r}_j\mapsto R\mathbf{r}_j$, this preserves the phase $(R\mathbf{k}_\alpha)\!\cdot\ (R\mathbf{r}_j)=\mathbf{k}_\alpha\!\cdot\!\mathbf{r}_j$, so the structure-factor computation is $SO(3)$-equivariant under global rotations.

After forward accumulation, we apply a learnable spectral filter to the structure factors in reciprocal space.
In our implementation, the linear reciprocal-space filter is shared across all degrees $\ell$.
Concretely, for each $\mathbf{k}_\alpha$ we use a single channel mixer
\begin{equation}
\widetilde{s}_{\alpha,m}^{(\ell)}=\mathbf{F}\,s_{\alpha,m}^{(\ell)},\qquad
\mathbf{F}\in\mathbb{R}^{C\times C},
\label{eq:pbc_shared_filter}
\end{equation}

where the same channel mixer $\mathbf F$ is shared across degrees $\ell$.
Crucially, for each fixed $\ell$, $\mathbf F$ is applied identically to all magnetic components $m=-\ell,\ldots,\ell$ (i.e., it mixes channels only),
thereby preserving $SO(3)$-equivariance.
We parameterize $\mathbf F$ in a low-rank (bottleneck) form
\begin{equation}
\mathbf{F}=\mathbf{W}_{\mathrm{up}}\mathbf{W}_{\mathrm{down}},
\qquad
\mathbf{W}_{\mathrm{down}}\in\mathbb{R}^{C_{\downarrow}\times C},\;\;
\mathbf{W}_{\mathrm{up}}\in\mathbb{R}^{C\times C_{\downarrow}}.
\end{equation}
Here $\mathbf W_{\mathrm{down}}\in\mathbb R^{C_{\downarrow}\times C}$ performs a down-projection from $C$ channels to a compact latent dimension $C_{\downarrow}$, and $\mathbf W_{\mathrm{up}}\in\mathbb R^{C\times C_{\downarrow}}$ performs the corresponding up-projection back to $C$ channels. This low-rank form reduces parameters and computation relative to a full $C\times C$ matrix while retaining an expressive linear spectral response. In implementation, $\mathbf W_{\mathrm{down}}$ and $\mathbf W_{\mathrm{up}}$ correspond to the weights of the down and up-projection linear layers, and their product is reused for all degrees $\ell$, whereas subsequent nonlinear updates are applied degree-wise.

\noindent \textbf{Aperiodic systems.} For aperiodic systems, reciprocal sampling is performed on a fixed Cartesian voxel grid instead of a cell induced
reciprocal lattice. Given a frequency resolution $\Delta k$ and cutoff $k_{\max}$, we enumerate integer triples
\begin{equation}
(n_x,n_y,n_z)\in\mathcal I_{\mathrm{box}}
=\{-N_k,\ldots,N_k\}^3,\qquad N_k=\left\lfloor \frac{k_{\max}}{\Delta k}\right\rfloor,
\end{equation}

\begin{equation}
\mathbf{k}_{n_x,n_y,n_z}=\Delta k\,(n_x,n_y,n_z),
\end{equation}
and retain only points inside the spherical cutoff $\|\mathbf{k}\|\le k_{\max}$.
We denote the retained set by $\{\mathbf{k}_\alpha\}_{\alpha=1}^{N_k}$. Unlike the periodic branch, we use a separable voxel window
\begin{equation}
D(\mathbf{k},\mathbf{r})
=\prod_{d\in\{x,y,z\}}
\operatorname{sinc}\!\left(\frac{\Delta k\, r_d}{2}\right),
\end{equation}
which reduces discretization artifacts from finite $\Delta k$. In practice, coordinates are first centered and mapped to an internal SVD frame before this damping is applied, improving numerical stability across differently oriented structures. Instead of a single $k$-independent mixer, each reciprocal point uses a radial embedding $\psi(\|\mathbf k_\alpha\|)$
and a bottleneck projection:
\begin{equation}
\mathbf f_\alpha
= \mathbf W_{\mathrm{up}}\!\left(\mathbf W_{\mathrm{down}}\,\psi(\|\mathbf k_\alpha\|)\right)\in\mathbb R^{C}.
\end{equation}
where $\psi(\|\mathbf k_\alpha\|)\in\mathbb R^{C_\psi}$ is a fixed radial embedding, 
$\mathbf W_{\mathrm{down}}\in\mathbb R^{C_\downarrow\times C_\psi}$ and 
$\mathbf W_{\mathrm{up}}\in\mathbb R^{C\times C_\downarrow}$ are learnable projections. Equivalently, this corresponds to using a diagonal channel mixer $\mathbf F(\mathbf k_\alpha)=\mathrm{diag}(\mathbf f_\alpha)$, a channel-wise spectral gate at each reciprocal point. Here, $\mathbf f_\alpha$ is applied channel-wise at $\mathbf k_\alpha$, giving a resolution-aware spectral response that varies with frequency magnitude. Since $\psi(\|\mathbf k_\alpha\|)$ depends only on the precomputed grid, these radial features can be prepared once and reused across structures, reducing runtime overhead.

\vspace{0.5em}
\noindent{\textbf{Inverse accumulation.}}
After reciprocal-space filtering, the long-range update is obtained in one step by degree-wise inverse accumulation, degree-specific nonlinear refinement, and scalar-conditioned gating:
\begin{equation}
\begin{aligned}
\mathbf{x}_i^{\mathrm{Ewald}}
&=\mathrm{Gate}\!\Bigg(
\mathbf{W}_g\,\mathbf{h}_{i,\ell=0},\;
\bigoplus_{\ell=0}^{\ell_{\max}}
\mathrm{MLP}^{(\ell)}\!\Big(
\eta\,[M_{m}^{(\ell)}(\mathbf r_i)]_{m=-\ell}^{\ell}
\Big)\Bigg),
\end{aligned}
\label{eq:ewald_from_M}
\end{equation}
where $\mathbf h_{i,\ell=0}\in\mathbb R^{C}$ denotes the scalar ($\ell=0$) block of the current atom representation and
$\mathbf g_i=\mathbf W_g\,\mathbf h_{i,\ell=0}\in\mathbb R^{\ell_{\max}C},
\mathbf W_g\in\mathbb R^{(\ell_{\max}C)\times C}
$
are the gating coefficients. In our implementation, $\mathbf g_i$ provides one $C$-dimensional gate per degree $\ell=1,\ldots,\ell_{\max}$, and the corresponding gate is broadcast
to all magnetic components $m=-\ell,\ldots,\ell$ within that degree.
Moreover, $M^{(\ell)}(\mathbf r_i)\in\mathbb R^{(2\ell+1)\times C}$ is the degree-$\ell$ inverse-accumulated message at atom $i$, and
$\mathrm{MLP}^{(\ell)}:\mathbb R^{(2\ell+1)\times C}\to\mathbb R^{(2\ell+1)\times C}$ is the degree-specific nonlinear update network that acts on channels and is shared across $m$.
Finally, $\mathrm{Gate}(\mathbf g_i,\cdot)$ denotes scalar-conditioned gating that modulates the non-scalar ($\ell>0$) blocks using $\mathbf g_i$.

\vspace{0.5em}\noindent \textbf{Information fusion.}
At interaction layer $t$, we fuse the running representation, the local update, and the long-range update via
\begin{equation}
\mathbf{x}_i^{t+1} = \frac{1}{\sqrt{3}}
\left(
\mathbf{x}_i^{t}
+ \mathbf{x}_i^{\mathrm{Local},t}
+ \mathbf{x}_i^{\mathrm{Ewald},t}
\right).
\end{equation}
Here, $t$ denotes the message-passing layer index. For brevity, earlier equations omit the layer superscript; all feature tensors can be interpreted as layer-dependent when this does not cause ambiguity.
% \vspace{0.5em}
% \noindent \textbf{Information Fusion.} The output of the chosen short-range backbone is fused with the long-range spectral representation via a unified residual update:
% \begin{equation}
%     \mathbf{x}_i^{l+1}
%     = \frac{1}{\sqrt{3}}
%     \left(
%         \mathbf{x}_i^{l}
%         + \mathbf{x}_i^{\text{Local},\,l}
%         + \mathbf{x}_i^{\text{Ewald},\,l}
%     \right)
% \end{equation}
% This ensures that the final atomic representation contains both high-frequency local information and low-frequency global structural information.

% For the eSCN-based model, atom-wise forces are derived via the negative gradient of the predicted potential energy with respect to atomic positions, utilizing automatic differentiation: 
% \begin{equation} 
%     \mathbf{F}i = -\nabla{\mathbf{r}_i} E_{\text{total}} 
% \end{equation} 
% For the Equiformer V2 model, we use a block of equivariant graph attention and treat the output of degree 1 as our forces predictions.

\vspace{0.5em}

\subsection{Free-energy calculation}
\label{sec:free_energy}

To characterize folding thermodynamics, we utilized the extensively sampled simulation trajectories from \cite{lindorff2011fast} (the same dataset as in \cite{wang2024ab}), comprising 100,000 snapshots for Chignolin. 
Snapshots were classified into folded and unfolded states using the native contact fraction $Q$ with thresholds of $Q > 0.82$ for folded and $Q < 0.03$ for unfolded states \cite{best2013native}. 
For each snapshot, the potential energy was re-evaluated using both eSCN and eSCN+EquiEwald potentials. 
State probabilities were then obtained by Boltzmann reweighting over the sampled conformational ensemble at temperature $T$,
\begin{equation}
P(s) \propto \sum_{i \in s} \exp\left(-\beta E_i\right),
\end{equation}
where $E_i$ denotes the re-evaluated potential energy of snapshot $i$, $\beta = (k_{\mathrm{B}}T)^{-1}$, and $s$ indicates either the folded or unfolded state. 
The free-energy difference between the two states was computed as
\begin{equation}
\Delta G = -k_{\mathrm{B}}T \ln \frac{P_{\mathrm{folded}}}{P_{\mathrm{unfolded}}}.
\end{equation}
This approach yields free-energy estimates directly from equilibrium ensembles without invoking additional biasing potentials or alchemical transformations.
% To characterize folding thermodynamics, simulation snapshots were first classified into folded and unfolded states using a structural order parameter (e.g., native contact fraction).
% Building upon the dataset and methodology of \cite{wang2024ab}, for each snapshot, the potential energy was re-evaluated using eSCN and eSCN + EquiEwald.
% State probabilities were then obtained by Boltzmann reweighting over the sampled conformational ensemble,
% \begin{equation}
% P(s) \propto \sum_{i \in s} \exp\left(-\beta E_i\right),
% \end{equation}
% where $E_i$ denotes the potential energy of snapshot $i$, $\beta = (k_{\mathrm{B}}T)^{-1}$, and $s$ indicates either the folded or unfolded state.

% The free-energy difference between the folded and unfolded states was computed as
% \begin{equation}
% \Delta G = -k_{\mathrm{B}}T \ln \frac{P_{\mathrm{folded}}}{P_{\mathrm{unfolded}}}.
% \end{equation}
% This procedure yields free-energy estimates directly from the equilibrium ensemble.

% % For temperature-dependent analyses, the same protocol was applied independently to trajectories generated at different temperatures, allowing the estimation of thermodynamic quantities such as melting temperatures and free-energy variations as a function of temperature.

\subsection{Training settings}
The model is optimized using a composite loss with energy $E$ and atom-wise forces $\mathbf{F}_i$:
\begin{equation}
    \mathcal{L}
    = \lambda_E \,\| E^{\text{pred}} - E^{\text{ref}} \|_1
    + \lambda_F \,\frac{1}{3N}\sum_{i=1}^{N} \| \mathbf{F}_i^{\text{pred}} - \mathbf{F}_i^{\text{ref}} \|_1 ,
\end{equation}
where $\|\cdot\|_1$ denotes the mean absolute error (MAE), 
$E^{\text{pred}}$ and $E^{\text{ref}}$ are the predicted and reference total energies per configuration, 
$\mathbf{F}_i^{\text{pred}},\mathbf{F}_i^{\text{ref}}\in\mathbb{R}^3$ are the predicted and reference forces on atom $i$, 
$N$ is the number of atoms in the configuration, and the factor $1/(3N)$ averages the force MAE over all $3N$ Cartesian components. 
$\lambda_E,\lambda_F>0$ are scalar weights balancing the two objectives.

The training settings on the OC20 periodic systems dataset are as follows: a cutoff radius of 6.0\AA\  is used to construct local neighborhood graphs. For reciprocal space settings, we followed~\cite{kosmala2023ewald}. In reciprocal space, the number of included frequencies along the three lattice directions is adjusted to account for the average anisotropy of the unit cell. Observing that the average reciprocal-space unit cell on OC20 is approximately three times narrower along the surface normal (z‑direction) than along the x and y directions, we set 1, 1, and 3 for each direction. In the loss function, the weighting coefficients for energy and forces are set to $\lambda_E = 1$ and $\lambda_F = 100$. 
For aperiodic systems, including the supramolecular buckyball catcher and the charged dimer, the following settings are employed. The spherical harmonic degree is set to $\ell_{\max} = 3$, and the Ewald-based long-range module uses a reciprocal-space cutoff of 0.6\AA\textsuperscript{-1}, a grid spacing of 0.2\AA\textsuperscript{-1}, and 128-dimensional Gaussian radial-basis functions to parameterize the frequency-domain filters. The loss-function weighting between energy and forces is fixed at $\lambda_E=1$ and $\lambda_F=100$. Additional experimental hyperparameters are provided in Supplementary Tables S3–S6.

\section*{Data availability}

The datasets used in this work are available at Code Ocean.

\section*{Code availability}
The source code for reproducing the findings in this paper is available at Code Ocean. It is licensed under the \texttt{Apache License 2.0}, which allows users to use, modify, and distribute the code freely, provided that proper attribution is given to the original authors. This open source approach improves the reproducibility of our results and facilitates further research in this area.

\bibliography{sn-bibliography}% common bib file

@article{hospital2015molecular,
  title={Molecular dynamics simulations: advances and applications},
  author={Hospital, Adam and Go{\~n}i, Josep Ramon and Orozco, Modesto and Gelp{\'\i}, Josep L},
  journal={Advances and Applications in Bioinformatics and Chemistry},
  pages={37--47},
  year={2015},
  publisher={Taylor \& Francis}
}

@article{senftle2016reaxff,
  title={The ReaxFF reactive force-field: development, applications and future directions},
  author={Senftle, Thomas P and Hong, Sungwook and Islam, Md Mahbubul and Kylasa, Sudhir B and Zheng, Yuanxia and Shin, Yun Kyung and Junkermeier, Chad and Engel-Herbert, Roman and Janik, Michael J and Aktulga, Hasan Metin and others},
  journal={npj Computational Materials},
  volume={2},
  number={1},
  pages={1--14},
  year={2016},
  publisher={Nature Publishing Group}
}

@article{karplus1990molecular,
  title={Molecular dynamics simulations in biology},
  author={Karplus, Martin and Petsko, Gregory A},
  journal={Nature},
  volume={347},
  number={6294},
  pages={631--639},
  year={1990},
  publisher={Nature Publishing Group UK London}
}

@article{yao2022applying,
  title={Applying classical, ab initio, and machine-learning molecular dynamics simulations to the liquid electrolyte for rechargeable batteries},
  author={Yao, Nan and Chen, Xiang and Fu, Zhong-Heng and Zhang, Qiang},
  journal={Chemical Reviews},
  volume={122},
  number={12},
  pages={10970--11021},
  year={2022},
  publisher={ACS Publications}
}

@article{geerlings2003conceptual,
  title={Conceptual density functional theory},
  author={Geerlings, Paul and De Proft, Frank and Langenaeker, Wilfried},
  journal={Chemical reviews},
  volume={103},
  number={5},
  pages={1793--1874},
  year={2003},
  publisher={ACS Publications}
}

@article{ramprasad2017machine,
  title={Machine learning in materials informatics: recent applications and prospects},
  author={Ramprasad, Rampi and Batra, Rohit and Pilania, Ghanshyam and Mannodi-Kanakkithodi, Arun and Kim, Chiho},
  journal={npj Computational Materials},
  volume={3},
  number={1},
  pages={54},
  year={2017},
  publisher={Nature Publishing Group UK London}
}

@article{butler2018machine,
  title={Machine learning for molecular and materials science},
  author={Butler, Keith T and Davies, Daniel W and Cartwright, Hugh and Isayev, Olexandr and Walsh, Aron},
  journal={Nature},
  volume={559},
  number={7715},
  pages={547--555},
  year={2018},
  publisher={Nature Publishing Group UK London}
}

@article{gubernatis2018machine,
  title={Machine learning in materials design and discovery: Examples from the present and suggestions for the future},
  author={Gubernatis, JE and Lookman, TJPRM},
  journal={Physical Review Materials},
  volume={2},
  number={12},
  pages={120301},
  year={2018},
  publisher={APS}
}

@article{batzner20223,
  title={E (3)-equivariant graph neural networks for data-efficient and accurate interatomic potentials},
  author={Batzner, Simon and Musaelian, Albert and Sun, Lixin and Geiger, Mario and Mailoa, Jonathan P and Kornbluth, Mordechai and Molinari, Nicola and Smidt, Tess E and Kozinsky, Boris},
  journal={Nature communications},
  volume={13},
  number={1},
  pages={2453},
  year={2022},
  publisher={Nature Publishing Group UK London}
}

@article{musaelian2023learning,
  title={Learning local equivariant representations for large-scale atomistic dynamics},
  author={Musaelian, Albert and Batzner, Simon and Johansson, Anders and Sun, Lixin and Owen, Cameron J and Kornbluth, Mordechai and Kozinsky, Boris},
  journal={Nature Communications},
  volume={14},
  number={1},
  pages={579},
  year={2023},
  publisher={Nature Publishing Group UK London}
}

@article{batatia2022mace,
  title={MACE: Higher order equivariant message passing neural networks for fast and accurate force fields},
  author={Batatia, Ilyes and Kovacs, David P and Simm, Gregor and Ortner, Christoph and Cs{\'a}nyi, G{\'a}bor},
  journal={Advances in neural information processing systems},
  volume={35},
  pages={11423--11436},
  year={2022}
}

@article{cheng2025latent,
  title={Latent Ewald summation for machine learning of long-range interactions},
  author={Cheng, Bingqing},
  journal={npj Computational Materials},
  volume={11},
  number={1},
  pages={80},
  year={2025},
  publisher={Nature Publishing Group UK London}
}

@inproceedings{kosmala2023ewald,
  title={Ewald-based long-range message passing for molecular graphs},
  author={Kosmala, Arthur and Gasteiger, Johannes and Gao, Nicholas and G{\"u}nnemann, Stephan},
  booktitle={International Conference on Machine Learning},
  pages={17544--17563},
  year={2023},
  organization={PMLR}
}

@article{schutt2018schnet,
  title={Schnet--a deep learning architecture for molecules and materials},
  author={Sch{\"u}tt, Kristof T and Sauceda, Huziel E and Kindermans, P-J and Tkatchenko, Alexandre and M{\"u}ller, K-R},
  journal={The Journal of chemical physics},
  volume={148},
  number={24},
  year={2018},
  publisher={AIP Publishing}
}

@article{anstine2023machine,
  title={Machine learning interatomic potentials and long-range physics},
  author={Anstine, Dylan M and Isayev, Olexandr},
  journal={The Journal of Physical Chemistry A},
  volume={127},
  number={11},
  pages={2417--2431},
  year={2023},
  publisher={ACS Publications}
}

@article{levin2009polarizable,
  title={Polarizable ions at interfaces},
  author={Levin, Yan},
  journal={Physical review letters},
  volume={102},
  number={14},
  pages={147803},
  year={2009},
  publisher={APS}
}

@article{bedrov2019molecular,
  title={Molecular dynamics simulations of ionic liquids and electrolytes using polarizable force fields},
  author={Bedrov, Dmitry and Piquemal, Jean-Philip and Borodin, Oleg and MacKerell Jr, Alexander D and Roux, Beno{\^\i}t and Schr{"o"}der, Christian},
  journal={Chemical reviews},
  volume={119},
  number={13},
  pages={7940--7995},
  year={2019},
  publisher={ACS Publications}
}

@article{borodin2009polarizable,
  title={Polarizable force field development and molecular dynamics simulations of ionic liquids},
  author={Borodin, Oleg},
  journal={The Journal of Physical Chemistry B},
  volume={113},
  number={33},
  pages={11463--11478},
  year={2009},
  publisher={ACS Publications}
}

@article{chanussot2021open,
  title={Open catalyst 2020 (OC20) dataset and community challenges},
  author={Chanussot, Lowik and Das, Abhishek and Goyal, Siddharth and Lavril, Thibaut and Shuaibi, Muhammed and Riviere, Morgane and Tran, Kevin and Heras-Domingo, Javier and Ho, Caleb and Hu, Weihua and others},
  journal={Acs Catalysis},
  volume={11},
  number={10},
  pages={6059--6072},
  year={2021},
  publisher={ACS Publications}
}

@article{chmiela2023accurate,
  title={Accurate global machine learning force fields for molecules with hundreds of atoms},
  author={Chmiela, Stefan and Vassilev-Galindo, Valentin and Unke, Oliver T and Kabylda, Adil and Sauceda, Huziel E and Tkatchenko, Alexandre and M{\"u}ller, Klaus-Robert},
  journal={Science Advances},
  volume={9},
  number={2},
  pages={eadf0873},
  year={2023},
  publisher={American Association for the Advancement of Science}
}

@article{king2025machine,
  title={Machine learning of charges and long-range interactions from energies and forces},
  author={King, Daniel S and Kim, Dongjin and Zhong, Peichen and Cheng, Bingqing},
  journal={Nature Communications},
  volume={16},
  number={1},
  pages={8763},
  year={2025},
  publisher={Nature Publishing Group UK London}
}

@article{wang2023aimd,
  title={AIMD-Chig: Exploring the conformational space of a 166-atom protein Chignolin with ab initio molecular dynamics},
  author={Wang, Tong and He, Xinheng and Li, Mingyu and Shao, Bin and Liu, Tie-Yan},
  journal={Scientific Data},
  volume={10},
  number={1},
  pages={549},
  year={2023},
  publisher={Nature Publishing Group UK London}
}

@article{best2013native,
  title={Native contacts determine protein folding mechanisms in atomistic simulations},
  author={Best, Robert B and Hummer, Gerhard and Eaton, William A},
  journal={Proceedings of the National Academy of Sciences},
  volume={110},
  number={44},
  pages={17874--17879},
  year={2013},
  publisher={National Academy of Sciences}
}

@article{lindorff2011fast,
  title={How fast-folding proteins fold},
  author={Lindorff-Larsen, Kresten and Piana, Stefano and Dror, Ron O and Shaw, David E},
  journal={Science},
  volume={334},
  number={6055},
  pages={517--520},
  year={2011},
  publisher={American Association for the Advancement of Science}
}

@article{wang2024ab,
  title={Ab initio characterization of protein molecular dynamics with AI2BMD},
  author={Wang, Tong and He, Xinheng and Li, Mingyu and Li, Yatao and Bi, Ran and Wang, Yusong and Cheng, Chaoran and Shen, Xiangzhen and Meng, Jiawei and Zhang, He and others},
  journal={Nature},
  volume={635},
  number={8040},
  pages={1019--1027},
  year={2024},
  publisher={Nature Publishing Group UK London}
}

@inproceedings{passaro2023reducing,
  title={Reducing SO (3) convolutions to SO (2) for efficient equivariant GNNs},
  author={Passaro, Saro and Zitnick, C Lawrence},
  booktitle={International conference on machine learning},
  pages={27420--27438},
  year={2023},
  organization={PMLR}
}

@article{liao2023equiformerv2,
  title={Equiformerv2: Improved equivariant transformer for scaling to higher-degree representations},
  author={Liao, Yi-Lun and Wood, Brandon and Das, Abhishek and Smidt, Tess},
  journal={arXiv preprint arXiv:2306.12059},
  year={2023}
}

@article{deringer2021gaussian,
  title={Gaussian process regression for materials and molecules},
  author={Deringer, Volker L and Bart{\'o}k, Albert P and Bernstein, Noam and Wilkins, David M and Ceriotti, Michele and Cs{\'a}nyi, G{\'a}bor},
  journal={Chemical reviews},
  volume={121},
  number={16},
  pages={10073--10141},
  year={2021},
  publisher={ACS Publications}
}

@article{thomas2018tensor,
  title={Tensor field networks: Rotation-and translation-equivariant neural networks for 3d point clouds},
  author={Thomas, Nathaniel and Smidt, Tess and Kearnes, Steven and Yang, Lusann and Li, Li and Kohlhoff, Kai and Riley, Patrick},
  journal={arXiv preprint arXiv:1802.08219},
  year={2018}
}

@article{huguenin2023physics,
  title={Physics-inspired equivariant descriptors of nonbonded interactions},
  author={Huguenin-Dumittan, Kevin K and Loche, Philip and Haoran, Ni and Ceriotti, Michele},
  journal={The Journal of Physical Chemistry Letters},
  volume={14},
  number={43},
  pages={9612--9618},
  year={2023},
  publisher={ACS Publications}
}

@article{zitnick2022spherical,
  title={Spherical channels for modeling atomic interactions},
  author={Zitnick, Larry and Das, Abhishek and Kolluru, Adeesh and Lan, Janice and Shuaibi, Muhammed and Sriram, Anuroop and Ulissi, Zachary and Wood, Brandon},
  journal={Advances in Neural Information Processing Systems},
  volume={35},
  pages={8054--8067},
  year={2022}
}

@article{gasteiger2020directional,
  title={Directional message passing for molecular graphs},
  author={Gasteiger, Johannes and Gro{\ss}, Janek and G{\"u}nnemann, Stephan},
  journal={arXiv preprint arXiv:2003.03123},
  year={2020}
}

@article{duval2023hitchhiker,
  title={A hitchhiker's guide to geometric gnns for 3d atomic systems},
  author={Duval, Alexandre and Mathis, Simon V and Joshi, Chaitanya K and Schmidt, Victor and Miret, Santiago and Malliaros, Fragkiskos D and Cohen, Taco and Lio, Pietro and Bengio, Yoshua and Bronstein, Michael},
  journal={arXiv preprint arXiv:2312.07511},
  year={2023}
}

@article{gao2025foundation,
  title={A foundation machine learning potential with polarizable long-range interactions for materials modelling},
  author={Gao, Rongzhi and Yam, ChiYung and Mao, Jianjun and Chen, Shuguang and Chen, GuanHua and Hu, Ziyang},
  journal={Nature Communications},
  volume={16},
  number={1},
  pages={10484},
  year={2025},
  publisher={Nature Publishing Group UK London}
}

@article{unke2021machine,
  title={Machine learning force fields},
  author={Unke, Oliver T and Chmiela, Stefan and Sauceda, Huziel E and Gastegger, Michael and Poltavsky, Igor and Schutt, Kristof T and Tkatchenko, Alexandre and Muller, Klaus-Robert},
  journal={Chemical Reviews},
  volume={121},
  number={16},
  pages={10142--10186},
  year={2021},
  publisher={ACS Publications}
}

@article{carleo2019machine,
  title={Machine learning and the physical sciences},
  author={Carleo, Giuseppe and Cirac, Ignacio and Cranmer, Kyle and Daudet, Laurent and Schuld, Maria and Tishby, Naftali and Vogt-Maranto, Leslie and Zdeborov{\'a}, Lenka},
  journal={Reviews of Modern Physics},
  volume={91},
  number={4},
  pages={045002},
  year={2019},
  publisher={APS}
}

@article{grisafi2019incorporating,
  title={Incorporating long-range physics in atomic-scale machine learning},
  author={Grisafi, Andrea and Ceriotti, Michele},
  journal={The Journal of chemical physics},
  volume={151},
  number={20},
  year={2019},
  publisher={AIP Publishing}
}

@article{geiger2022e3nn,
  title={e3nn: Euclidean neural networks},
  author={Geiger, Mario and Smidt, Tess},
  journal={arXiv preprint arXiv:2207.09453},
  year={2022}
}

@article{weiler20183d,
  title={3d steerable cnns: Learning rotationally equivariant features in volumetric data},
  author={Weiler, Maurice and Geiger, Mario and Welling, Max and Boomsma, Wouter and Cohen, Taco S},
  journal={Advances in Neural information processing systems},
  volume={31},
  year={2018}
}

@article{chmiela2017machine,
  title={Machine learning of accurate energy-conserving molecular force fields},
  author={Chmiela, Stefan and Tkatchenko, Alexandre and Sauceda, Huziel E and Poltavsky, Igor and Sch{\"u}tt, Kristof T and M{\"u}ller, Klaus-Robert},
  journal={Science advances},
  volume={3},
  number={5},
  pages={e1603015},
  year={2017},
  publisher={American Association for the Advancement of Science}
}

@article{niblett2021learning,
  title={Learning intermolecular forces at liquid--vapor interfaces},
  author={Niblett, Samuel P and Galib, Mirza and Limmer, David T},
  journal={The Journal of chemical physics},
  volume={155},
  number={16},
  year={2021},
  publisher={AIP Publishing}
}

@article{unke2021spookynet,
  title={SpookyNet: Learning force fields with electronic degrees of freedom and nonlocal effects},
  author={Unke, Oliver T and Chmiela, Stefan and Gastegger, Michael and Sch{\"u}tt, Kristof T and Sauceda, Huziel E and M{\"u}ller, Klaus-Robert},
  journal={Nature communications},
  volume={12},
  number={1},
  pages={7273},
  year={2021},
  publisher={Nature Publishing Group UK London}
}

@article{unke2019physnet,
  title={PhysNet: A neural network for predicting energies, forces, dipole moments, and partial charges},
  author={Unke, Oliver T and Meuwly, Markus},
  journal={Journal of chemical theory and computation},
  volume={15},
  number={6},
  pages={3678--3693},
  year={2019},
  publisher={ACS Publications}
}

@article{ko2021fourth,
  title={A fourth-generation high-dimensional neural network potential with accurate electrostatics including non-local charge transfer},
  author={Ko, Tsz Wai and Finkler, Jonas A and Goedecker, Stefan and Behler, J{\"o}rg},
  journal={Nature communications},
  volume={12},
  number={1},
  pages={398},
  year={2021},
  publisher={Nature Publishing Group UK London}
}

@article{gao2022self,
  title={Self-consistent determination of long-range electrostatics in neural network potentials},
  author={Gao, Ang and Remsing, Richard C},
  journal={Nature communications},
  volume={13},
  number={1},
  pages={1572},
  year={2022},
  publisher={Nature Publishing Group UK London}
}

@article{shaidu2024incorporating,
  title={Incorporating long-range electrostatics in neural network potentials via variational charge equilibration from shortsighted ingredients},
  author={Shaidu, Yusuf and Pellegrini, Franco and K{\"u}{\c{c}}{\"u}kbenli, Emine and Lot, Ruggero and de Gironcoli, Stefano},
  journal={npj Computational Materials},
  volume={10},
  number={1},
  pages={47},
  year={2024},
  publisher={Nature Publishing Group UK London}
}

@article{zhang2022deep,
  title={A deep potential model with long-range electrostatic interactions},
  author={Zhang, Linfeng and Wang, Han and Muniz, Maria Carolina and Panagiotopoulos, Athanassios Z and Car, Roberto and others},
  journal={The Journal of Chemical Physics},
  volume={156},
  number={12},
  year={2022},
  publisher={AIP Publishing}
}

@article{schutt2017schnet,
  title={Schnet: A continuous-filter convolutional neural network for modeling quantum interactions},
  author={Sch{\"u}tt, Kristof and Kindermans, Pieter-Jan and Sauceda Felix, Huziel Enoc and Chmiela, Stefan and Tkatchenko, Alexandre and M{\"u}ller, Klaus-Robert},
  journal={Advances in neural information processing systems},
  volume={30},
  year={2017}
}

@article{deng2023chgnet,
  title={CHGNet as a pretrained universal neural network potential for charge-informed atomistic modelling},
  author={Deng, Bowen and Zhong, Peichen and Jun, KyuJung and Riebesell, Janosh and Han, Kevin and Bartel, Christopher J and Ceder, Gerbrand},
  journal={Nature Machine Intelligence},
  volume={5},
  number={9},
  pages={1031--1041},
  year={2023},
  publisher={Nature Publishing Group UK London}
}

@article{yu2022capturing,
  title={Capturing long-range interaction with reciprocal space neural network},
  author={Yu, Hongyu and Hong, Liangliang and Chen, Shiyou and Gong, Xingao and Xiang, Hongjun},
  journal={arXiv preprint arXiv:2211.16684},
  year={2022}
}

\end{document}